\documentclass[11pt]{article}
\usepackage[utf8]{inputenc}
\usepackage[T1]{fontenc} 
\usepackage[english]{babel} 

\usepackage[left=2.5cm,right=2.5cm,top=3cm,bottom=3cm]{geometry}

\usepackage{amsmath}
\usepackage{amsthm}
\usepackage{amssymb}
\usepackage[normalem]{ulem}

\usepackage{authblk}

\usepackage{enumitem}

\usepackage{dsfont}

\usepackage{graphicx}
\usepackage{color}

\usepackage[pdftex,%
colorlinks=true,linkcolor=blue,citecolor=blue,urlcolor=blue
]
{hyperref}

\usepackage{comment}
\usepackage{ stmaryrd }
\usepackage{physics}

\theoremstyle{plain}
\newtheorem{Theorem}{Theorem}
\newtheorem{Lemma}{Lemma}
\newtheorem{prop}{Proposition}
\newtheorem{cor}{Corollary}

\theoremstyle{definition}
\newtheorem{Definition}{Definition}
\newtheorem{hyp}{Assumption}
\newtheorem{rk}{Remark}
\newtheorem{ex}{Example}

\newtheorem*{Proof}{Proof}

\newcommand{\HH}{\mathcal H}

\definecolor{dark-green}{RGB}{0, 128, 0}

\DeclareMathOperator{\Sp}{Sp}

\DeclareMathOperator{\Spec}{Spec}

    \title{Estimating bulk and edge topological indices\\ in finite open chiral chains}
\author[1]{Lucien Jezequel}
\author[2]{Clément Tauber}
\author[3]{Pierre Delplace}
\affil[1,3] {ENSL, CNRS, Laboratoire de physique, F-69342 Lyon, France.}
\affil[2]{Institut de Recherche Mathématique Avancée, UMR 7501 Université de Strasbourg et CNRS, 7 rue René-Descartes, 67000 Strasbourg, France}
\date{\today}

\begin{document}

\maketitle

\begin{abstract}
We develop a formalism to extend, simultaneously, the usual definition of bulk and edge indices from topological insulators to the case of a finite sample with open boundary conditions, and provide a physical interpretation of these quantities. We then show that they converge exponentially fast to an integer value when we increase the system size, and also that bulk and edge quantities coincide at finite size.
%
%
The theorem applies to any non-homogeneous system such as disordered or defect configurations. We focus on one-dimensional chains with chiral symmetry, such as the Su-Schrieffer-Heeger model, but the proof actually only requires the Hamiltonian  to be short-range and with a spectral gap in the bulk. The definition of bulk and edge indices relies on a finite-size version of the switch-function formalism where the Fermi projector is smoothed in energy using a carefully chosen regularization parameter.
\end{abstract}

\section{Introduction}

Topological insulators are a special class of materials that are gapped in their bulk but exhibit edge modes at the Fermi energy. If they have been first discovered in the quantum Hall effect of two-dimensional lattices \cite{Klitzing, Thouless}, topological insulators are actually found in all dimensions \cite{Kitaev2009,Ryu_2010,hasan2010colloquium,ProdanEmilSchultz}. What characterizes those materials is that their number of edge modes is a topological quantity which is invariant as long as the material has a bulk gap. In fact there is a staggering relation, called the bulk-edge correspondence, which relates this quantity to another topological index defined in the bulk \cite{HatsugaiPRL1993}. 

The topological nature of the edge modes is associated to a remarkable stability against physical perturbations and relies on the definition of indices which involve various areas of mathematics. Bulk and edge indices, as well as their correspondence, have been studied for translation-invariant system using fiber bundle theory \cite{graf2013bulk} as well as disordered systems using non-commutative geometry, K-theory, and Fredholm theory \cite{bellissard1994noncommutative,avron1994charge,Kellendonk}. All indices share the common feature of being usually well defined mathematically for infinite (bulk) or half-infinite (edge) systems. In particular, such indices trivially vanish when applied to finite samples with open boundary conditions. This fact is problematic from the experimental point of view but also for numerical purposes to actually compute edge indices on a finite sample. In the last decades, several strategy have been followed to compute a numerical estimate of bulk or edge indices in finite samples \cite{bianco2011mapping,mondragon2014topological,prodan2017computational,loring2017finite,Localizer,tauber2018effective,michala2021wave}. 


In this paper we develop a formalism to extend, in a meaningful way, the definition of bulk and edge indices to the case of finite open chains. In general, these looking-like-index quantities  are not exactly quantized at finite size, but we show that they converge exponentially fast towards the same integer when the size of the system is increased. We also show that bulk and edge quantities coincide at finite size. Our main theorem applies to one-dimensional chains with chiral symmetry, such as the celebrated Su-Schrieffer-Heeger model \cite{SSH}, but in principle all the arguments could be extended to higher dimensional systems or other symmetry classes. The main point is that, for any short-range couplings, bulk and edge quantities are localized in distinct regions of space which are well separated for large enough chains, so that they can both be computed in the same system.

Beyond its ability to estimate both indices on the same sample, our approach is deterministic and applies to any non-homogeneous situation such as disordered configurations, defects or domain walls \cite{NonlinEdgeMode1D}, as long as the bulk spectral gap remains open. We illustrate this point on a numerical example. Moreover, our approach is reminiscent from the switch-function formalism \cite{avron1994charge,tauber2018effective}, but our proof of the main theorem only deals with finite matrices and does not directly refer to operators or infinite-dimensional indices, and hence bypasses any proof of trace-class property. As a byproduct, we believe our approach to be rather accessible. Finally, the central idea of this approach is to work with regularized (smooth) spectral functions instead of discontinuous ones like the Fermi projection. The regularization parameter is analog to a temperature and has to be carefully chosen for the indices to be almost quantized. 


The article is organized as follows. We describe our setting and discuss the main results in Section 2, together with a numerical example. Section 3 proves the main theorem by relying on some intermediate results about localization in space of bulk and edge quantities. Section 4 proves the latter results.

\section{Setting and main results}

\subsection{One-dimensional chiral chain}
We consider tight-binding  models on finite open chains. In the single-particle picture, the Hilbert space is $\HH_L = \ell^2(\llbracket 0,L-1 \rrbracket)\otimes \mathbb C^2$ where $L$ is the length of the chain and $\mathbb C^2$ stands for two internal degrees of freedom that we denote by $A$ and $B$. We consider a Hamiltonian $H$ acting on $\HH_L$, namely a Hermitian matrix of size $2L$, and we assume it has the chiral symmetry:
\begin{equation}\label{chiral_symmetry}
HC+CH=0, \qquad C = \begin{pmatrix}\mathds{1}_A&0\\0&-\mathds{1}_B\end{pmatrix},
\end{equation}
with $C$ the chiral operator.  Typical models we have in mind are, among many, the celebrated SSH chain \cite{SSH}, where $A$ and $B$ stand for distinct sublattice sites, and the Shockley model where $A$ and $B$ stand for distinct orbitals on every sites \cite{Shockley39}. The generalization to  models with even internal degrees of freedom is straightforward by replacing $\mathbb C^2$ by $\mathbb C^{2m}$ for $m \in \mathbb N$.


We denote by $\ket{x,s}$ the canonical basis of $\HH_L$, with $x \in \llbracket 0,L-1 \rrbracket$ and $s \in \{A,B\}$. The matrix elements of $H$ are $H_{x,x'}^{s,s'}=\bra{x,s} H \ket{x',s'}$. The chiral symmetry \eqref{chiral_symmetry} implies that $H_{x,x'}^{A,A} = H_{x,x'}^{B,B} = 0$, so that  $H$ is off-diagonal in the $(A,B)$-basis. 


Two central assumptions are required for the theorem below to apply: the model has to be short range, and the corresponding bulk Hamiltonian must have a spectral gap. In order to formulate them properly, it is convenient to consider a bulk extension of $H$, denoted by $H^{\text{bulk}}$, which acts on the infinite chain Hilbert space $\HH^\mathrm{bulk}=\ell^2(\mathbb Z) \otimes \mathbb C^2$ so that $H = \iota^* H^{\text{bulk}} \iota$ with $\iota$ the canonical injection of $\HH_L$  into $\HH^\mathrm{bulk}$ and $\iota^*$ is meanwhile the canonical truncation of $\HH^\mathrm{bulk}$ to $\HH_L$\footnote{Explicitly, one has for $\varphi \in \HH_L$ and $\varphi^\mathrm{bulk} \in \HH^\mathrm{bulk}$
$$(\iota \varphi)_x = \left\lbrace\begin{array}{ll}
    \varphi_x, & x \in \llbracket 0, L-1 \rrbracket, \\
    0, & \mathrm{otherwise},
\end{array}\right. \qquad  
(\iota^* \varphi^\mathrm{bulk})_x = \begin{array}{ll}
    \varphi^\mathrm{bulk}_x, & x \in \llbracket 0, L-1 \rrbracket.
\end{array}$$}.

This is nothing but saying that $H$ is the restriction of $H^{\text{bulk}}$ on $\HH_L$ with open boundary conditions. Notice that local perturbations at the boundary would not change the result, see Remark~\ref{rk:translationinvariance} below.

\begin{hyp}\label{hyp:shortrange}
The bulk Hamiltonian is short range: there exists a characteristic length $d>0$ and some constant $K_d>0$ such that
		$$
		\sup_{x\in \mathbb Z} \sum_{x' \in \mathbb Z} \norm{H^{\text{bulk}}_{x,x'}}e^{|x-x'|/d}\leq K_d < +\infty,
		$$
where $H^{\text{bulk}}_{x,y}$ are the matrix elements of $H^{\text{bulk}}$ in the canonical basis of $\ell^2(\mathbb Z^d)$ and $\norm{\cdot}$ is the operator norm on finite matrices.
\end{hyp}
This assumption is very mild and covers most of the tight-binding models from the literature: any finite range Hamiltonian,
$$H_{x,x'}^\mathrm{bulk} = 0, \qquad |x-x'|>r,$$
such as nearest-neighbor hopping ($r=1$) trivially satisfies it for any $d$. The assumption also allows Hamiltonians with exponentially decaying matrix elements 
$$\|H_{x,y}^\mathrm{bulk}\| \leq Ce^{-|x-x'|/\Tilde{d}},$$
as long as $\Tilde{d}<d$.  The main category of excluded physical models  are those with long range interaction (e.g. slow algebraic decay) where bulk and edge components are usually coupled and cannot be separated. Continuous models are also excluded as they have an infinite density of degree of freedom and would need another cut-off in the UV limit to obtain finite quantities. Finally, notice that Assumption \ref{hyp:shortrange} implies the same inequality for $H$ on the open chain, with the important point that $d$ and $K_d$ are independent of $L$.

\begin{hyp}\label{hyp:gap}
	The bulk Hamiltonian $H^{\text{bulk}}$ has a spectral gap around zero: there exists $\Delta>0$ such that $\Spec(H^{\text{bulk}}) \cap  (-\Delta,\Delta) = \emptyset$. 
\end{hyp}
Another equivalent formulation of this assumption would be that the restriction of $H^\mathrm{bulk}$ to $\HH_L$ with periodic boundary conditions has no eigenvalues in the interval $(-\Delta,\Delta)$, with $\Delta$ independent of $L$. However this does not imply the existence of a gap for $H$ with open boundary conditions, due to the presence of edge modes near zero energy.

\begin{figure}[b]
    \centering
    \includegraphics[width=15cm]{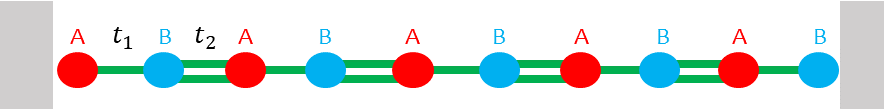}
    \caption{Chain of atoms in a SSH model}
    \label{fig:SSHchain}
\end{figure}

\begin{ex}\textit{SSH chain:}\label{ex:SSH} A typical chiral chain that exhibits topological properties is the SSH chain \cite{SSH}. It consists of alternating atoms of type A and B which are coupled by two alternating coefficients $t_{1,x}$ and $t_{2,x}$ (see Fig. \ref{fig:SSHchain}). The matrix elements read, for $x\in \mathbb Z^d$
\begin{align}
H^\mathrm{bulk}_{x,x}=\begin{pmatrix}0 & t_{1,x}\\t_{1,x} & 0\end{pmatrix}, \qquad H^\mathrm{bulk}_{x,x+1}=\begin{pmatrix}0 & t_{2,x}\\t_{2,x} & 0\end{pmatrix} = H^\mathrm{bulk}_{x+1,x} 
\label{eq:SSH}
\end{align}
and 0 otherwise, where $t_{1,x}$ an$_{2,x}$ and usually taken as constant $t_1$ and $t_2$ (homogeneous case).
As the interactions are of finite range, this model satisfies Assumption 1, e.g. with $d=1$ and $K_d = (t_1+t_2)e$. Moreover, one can check that the spectrum of this bulk Hamiltonian in the homogeneous case has a gap of size $2\Delta =2\left||t_1|- |t_2|\right|$ when $|t_1|\neq |t_2|$, and hence satisfies  Assumption 2. This model is known to be trivial in the case where $|t_1|> |t_2|$ and to have non-trivial topological properties when $|t_1|< |t_2|$ \cite{DelplaceZakphase,asboth2016schrieffer}.
\end{ex}

\begin{rk}
\textit{Beyond translation invariance:}\label{rk:translationinvariance}
We stress that no further assumption is required neither on $H$ nor on $H^{\text{bulk}}$. In particular, the model does not have to be translation invariant. The result below applies to any disordered or non-homogeneous configuration that is short-range, preserves the chiral symmetry and the bulk spectral gap. It is not required for the disorder to be ergodic and no average over disorder is required. Therefore, inhomogeneities like point defects or domain walls are also naturally taken into account. As a byproduct, adding a small potential $V$ supported near the edges of the chain allows one to consider a wide family of boundary conditions.
\end{rk}


\subsection{Bulk and edge indices}

In the following, we speak about indices with a slight abuse of language: The quantities that we consider are defined on finite Hilbert spaces, and thus are not strictly quantized. By the term index, we actually mean finite-size estimate which is almost quantized in the large size limit.

\medskip

In order to define indices, we would like to have an operator which flattens the bands like the Fermi projector. But we also want an operator which is short-range, in order to separate bulk and edge contributions. We will show latter that one way of doing so is to consider smooth operators in energy. Therefore, we introduce the operator
$$
S= \tanh(\frac{1}{\delta} H)
$$ 
defined for $\delta >0$, and which has the same eigenstates as $H$ but with eigenvalues changed from $E$ to $\tanh(E/\delta)$. $\delta$ is a regularization parameter and $S$ is a regularized version of the operator sign$(H)$ (We prefer to work with the operator sign$(H)$ instead of the usual Fermi projection $P = (\text{sign}(H)+1))/2$ in order to have slightly easier computations).

For $\delta \ll \Delta$, $S$ flattens the spectrum and discriminates between the upper band and lower band of $H$. Meanwhile, $1-S^2$ filters out all the bulk bands, leaving us with only the edge states inside the gap, if any.
Moreover, this operator can be shown to be short-range: its matrix elements decay exponentially, like in Assumption \ref{hyp:shortrange}, but with a rescaled distance $d'\sim dK_d/\delta$. This property is true even if $H$ is gapless, see Proposition~\ref{prop:combes-thomas} below.



Finally, we shall also filter in space by considering a step function $\theta : \llbracket 0,L-1 \rrbracket \to \mathbb R$ which is 0 near the right edge $(x=L-1)$, 1 near the left edge $(x=0)$, and jumps from 1 to 0 in the middle of the chain. We denote by $\theta(X)$ the multiplicative operator associated to $\theta$, namely $(\theta(X) \psi)_x = \theta(x) \psi_x$ for $\psi \in \HH_L$.

\begin{Definition}\label{def:indices}
The bulk index is defined by
\begin{equation}
 \mathcal{I}_{\mathrm{bulk}} = \frac{1}{2}\Tr\left(CS[\theta(X),S]\right), 
\end{equation}
and the edge index is defined by 
\begin{equation}
    \mathcal{I}_{\mathrm{edge}} = \Tr\left(C\theta(X)(1-S^2)\right).
\end{equation}
\end{Definition}
Up to the regularisation by $\delta$, the bulk index expression is analog to the one for the infinite and disordered chiral chain \cite[Eq.~(2.6)]{graf_bulk-edge_2018}, which itself can be reduced to the usual winding number for translation invariant systems \cite{DelplaceZakphase,asboth2016schrieffer}. It is remarkable that this regularisation procedure generalises the formula to  finite open chains and allows one to capture both bulk and edge indices in the same chain.

The edge index expression has a direct physical interpretation. Consider an eigenbasis $\psi_\lambda$ common to $C$ and $H^2$ (possible since $[C,H^2]=0$) with eigenvalues $C_\lambda$ and $E_\lambda^2$. We have
\begin{equation}
    \mathcal{I}_{\mathrm{edge}} = \sum_{\lambda} C_\lambda (1-\tanh(|E_\lambda|/\delta)^2) \sum_x \theta(x) |\psi_\lambda(x)|^2
\end{equation}
The term $\sum_x \theta(x) |\psi_\lambda(x)|^2$ corresponds to an integrated density of states in the region where the step function $\theta$ is 1, namely in the left half of the chain. The term $(1-\tanh(|E_\lambda|/\delta)^2)$ filters this density of states near zero energy  \cite{loring2021locality}. %
Finally the term $C_\lambda$ add a sign depending of the chiral charge of the state. Thus, the edge index is the chiral density of low energy states integrated in the left part of the chain.  If the bulk Hamiltonian has a gap $(-\Delta,\Delta)$ and $\delta \ll \Delta$, the edge index counts the  polarization of the edge modes near zero energy, and localized in the left part of the chain \cite{KaneLubensky,guzman2021geometry}.

\begin{Theorem}\label{thm:main}
	Let $H$ be a chiral Hamiltonian such that $H^\mathrm{bulk}$ satisfies Assumptions~\ref{hyp:shortrange} and \ref{hyp:gap}. Then, if we set $\delta =\sqrt{\frac{128\Delta dK_d}{ L}}$ we have:
	\begin{equation}
		\mathcal{I}_{\mathrm{edge/bulk}} \in \mathds{Z} + \gamma e^{-\sqrt{\frac{ L\Delta}{128d K_d }}}
		\label{majoration}
	\end{equation}
	where $\gamma$ is dominated by a polynomial function in the variables $(K_d,d,L)$.
\end{Theorem}

The bulk and edge indices from Definition~\ref{def:indices} are defined on a finite chain and are not expected to be exactly quantized. However, this theorem shows that they are quantized up to an error which decays exponentially as 
$\sqrt{\frac{ L\Delta}{128d K_d }}$.
More precisely, the physical regime to consider is $\Delta L \gg dK_d$ and take $\delta \sim \sqrt{\frac{\Delta dK_d}{ L}}$ where $\Delta$ is half the size of the bulk gap, $L$ is the size of the chain, $d$ is the coupling range and $K_d$ is the characteristic coupling strength. Consequently, in such a regime, the finite-size indices become able to discriminate between two distinct topological phases as soon as the error is smaller than $1/2$.

It should be noted that the condition $\Delta L \gg dK_d$ is, in some sense, equivalent to the quantisation condition which have already been determined in the case of (fully gapped) finite chains with periodic boundary conditions \cite{Toniolo}. The main addition to deal with the open boundary condition case (and its gapless edge states) is therefore this regularisation parameter $\delta$. 

\begin{figure}[t]
    \centering
    \includegraphics[width =14cm]{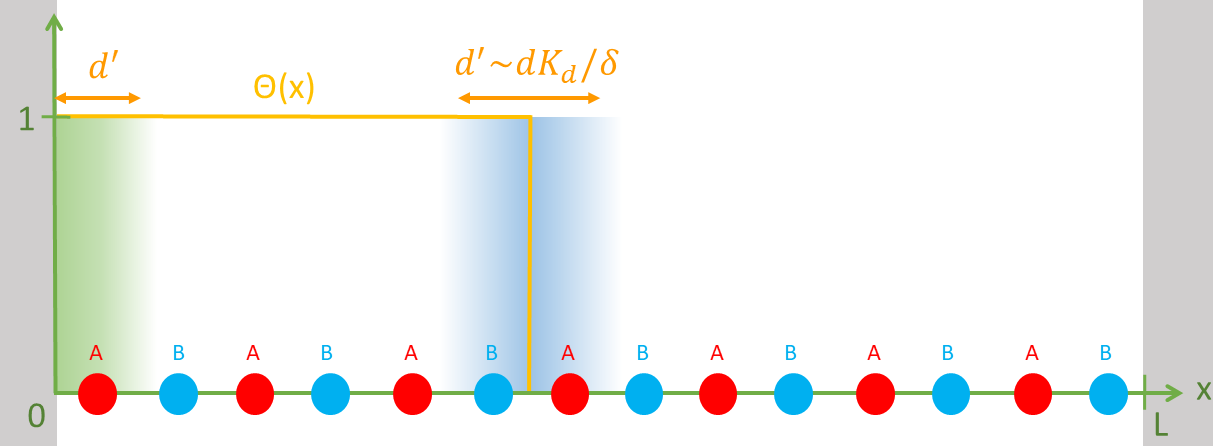}
    \caption{Chain of atoms with a chiral partition in two groups of sites A and B. For a given step function $\theta$ the edge/bulk index mainly depends on the coefficients localised in the green/blue region of characteristic length $d' \sim d K_d/\delta$}
    \label{Schema}
\end{figure}

Another important property of these indices is the localization in space of the operators involved. In order to compute the edge index, we do not need to know $H$ everywhere but only near the edge (green region in Fig \ref{Schema}). For the bulk index, all the information is localized near the transition of $\theta$ (blue region in Fig \ref{Schema}). The characteristic length of these regions is $d' = d K_d/\delta$, which appears all along the proof of Theorem~\ref{thm:main}. 

In particular, the distance between the edge and the transition of $\theta$ must be larger than $d'$, but it is common to consider a transition at $L/2$ for $L$ large enough. Moreover, if we truncate the system at some finite size much larger than $d'$, we only make an exponentially small error. This property is useful for numerical studies as it can reduce considerably the computation time but also indirectly for experiments as it means that local probes can access the indices. A similar property is used in \cite{bachmann2020many} to define bulk indices in interacting two-dimensional systems.

Notice that the exponential decay with $L$ and the localization properties are also illustrated in a numerical example, see Figure~\ref{fig:numerical_edge} below.

Finally we also show that the bulk and edge indices are in fact equal at finite size via the following correspondence:
\begin{prop}\label{prop:BEC}
Let $H$ be a chiral Hamiltonian on $\HH_L$. Then for any $\delta>0$ and any $L \in \mathbb N$ we have
$$
\mathcal{I}_{\mathrm{bulk}}= \mathcal{I}_{\mathrm{edge}}.
$$
\end{prop}
This result looks like a bulk-edge correspondence at finite size, except that the indices are actually not quantized. This allows us to reduce the proof of Theorem~\ref{thm:main} by focusing on the edge index only. The proof of this proposition is a few lines of algebra, see Section~\ref{sec:proofBEC} below.



\begin{figure}[htb]
    \centering
    \includegraphics[height=4cm,width=16cm]{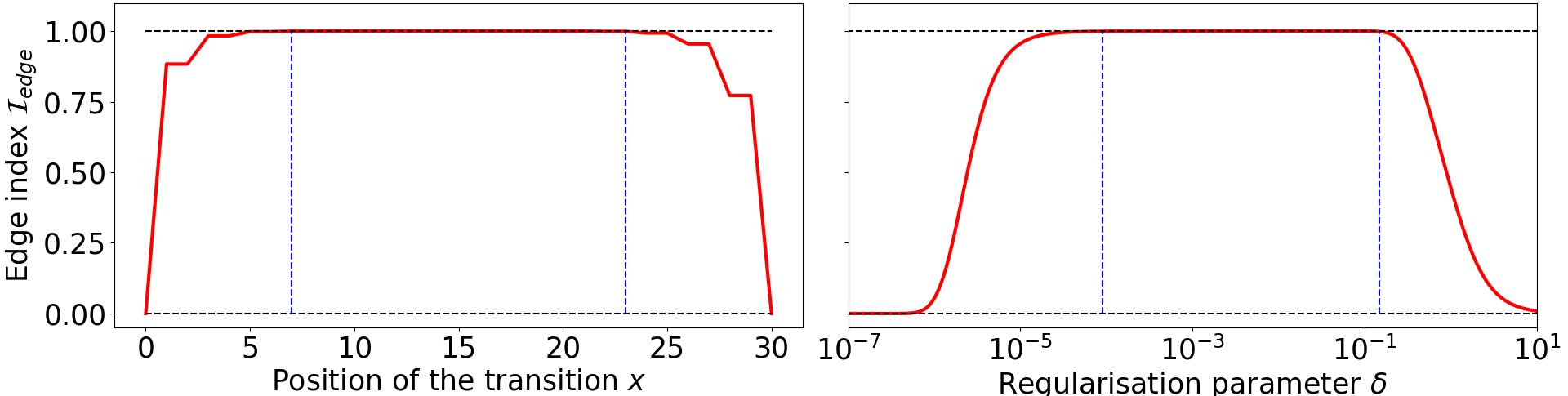}
    \caption{Left: Influence on $\mathcal I_\mathrm{edge}$ of the position of the transition for $\theta(X)$ ($L=30,\,\delta=1/20$). Right: Influence of $\delta$ ($L=30$).}
    \label{fig:numerical_parameters}
\end{figure}

\begin{rk}\textit{Freedom on $\delta$}:\label{rk:delta}
	The choice of $\delta$ in Theorem~\ref{thm:main} is the result of a trade-off between two facts. On the one hand, we need $\delta \ll \Delta$ so that $1-S^2$ only selects the states living in the spectral gap of the bulk Hamiltonian $H^{\text{bulk}}$, namely the edge modes. On the other hand, the edge modes on each side of the finite chain are always weakly coupled to each other and hence never exactly have  zero energy. Thus, they would be missed in the edge index by taking $\delta=0$ as it is done in half-infinite systems \cite{graf_bulk-edge_2018,KaneLubensky}. Therefore, we want $\delta$ small but not too small. This trade-off is reminiscent of the Robertson uncertainty relation $\Delta H \Delta X \geq |\langle [H,X]\rangle|/2$ \cite{UncertaintyRobertson} which relates the minimal uncertainty in energy $\Delta H$ and in position $\Delta X$ to the commutator $[H,X]$. The latter is proportional to the characteristic inter-site couplings times their distance. Here, one has $\Delta H \sim \delta$ and $\Delta X \sim d'\sim dK_d/\delta$, the characteristic correlation distance of $S$. Therefore, $\delta d' = K_dd$  looks similar, in the scaling, to the uncertainty relation. The proof of the theorem shows that when $L \gg \Delta X \sim d'\sim dK_d/\delta$ and $\Delta \gg \Delta H \sim \delta$, the indices become quantized in good approximation. This trade-off is illustrated by varying $\delta$ in a numerical example, see Figure \ref{fig:numerical_parameters} (right) below.
	
	Furthermore, notice that $S$ can be written $S = 2P_\delta -1$ where $P_\delta = 1/\left(1+e^{H/\delta}\right)$ is the thermal state associated to $H$ at temperature $T=\delta/k_B$. Therefore, as in condensed matter experiments, the temperature is small but never zero, this trade-off may naturally be satisfied as the thermal energy is often much smaller than the size of the gap, whereas the size of the sample is often much larger than the typical thermal correlation length. See also some recent mathematical work about extending two-dimensional bulk and edge quantities to (physical) finite temperature \cite{cornean2021general}.
	
	
\end{rk}

\begin{rk}\textit{Peculiarity of SSH chains:}\label{rk:SSH}
When $A$ and $B$ refer to distinct lattice sites, like in the SSH model, it is also possible to work with the Hilbert space $\ell^2(\llbracket 0,L-1\rrbracket)$ or $\ell^2(\llbracket 0,L\rrbracket)$ instead of $\HH_L$, with each site being alternatively $A$ or $B$, e.g. $A$ for the odd sites and $B$ for the even ones. In that case, there might be a mismatch between the end of the chain and the jump of the step function $\theta$: they could occur on a distinct type of site. In that case, the equality in Proposition~\ref{prop:BEC} must be replaced by
\begin{equation}
	\mathcal{I}_{\mathrm{edge}} = \mathcal{I}_{\mathrm{bulk}} +n_{A,\theta}-n_{B,\theta} \label{bulk-edge imbalanced}
\end{equation}
where $n_{A,\theta}-n_{B,\theta}$ counts the difference of number of $A/B$ site in the region where $\theta =1$. It can be interpreted as a chiral polarization of the sites in the support of $\theta$ and implies that the number of edge modes do not entirely depend on pure-bulk properties, as already pointed in \cite{KaneLubensky,guzman2021geometry}. This extra term is $H$-independent and  a pure lattice property, therefore it is still easy to compute. Often, we can make the choice to work with $\HH_L$ above giving a consistent choice of unit cell for the edges and $\theta$, so that this extra term is always zero in Proposition~\ref{prop:BEC}.

It is also possible to consider systems with odd degrees of freedom or having an imbalance of A/B sites per unit cell. However these model would violate the Assumption \ref{hyp:gap} by having zero energy bulk bands. This can be seen by contradiction using the relation \eqref{bulk-edge imbalanced}, as the  edge and bulk indices should be bounded whereas $n_{A,\theta}-n_{B,\theta}$ would increase linearly with the distance of the cut-off from the edge.
\end{rk}

\begin{rk}\textit{Higher dimensional indices:}\label{rk:generalisation} 
For the sake of the clarity, we focused this paper on implementing 1D chiral indices of finite chains. For this we highlight the importance of regularising  the usual Fermi projection on a energy scale $\delta$ which should be carefully chosen. This analysis is performed using some functional calculus (propositions \ref{prop:combes-thomas} and \ref{prop:edge-loc}) which are general and could be adapted to higher dimensional lattices. Therefore we strongly believe that the regularisation process is a key element which could also be used to define other bulk and edge $\mathds{Z}$-indices in (non-homogeneous) finite open systems, such as Chern or Floquet insluator invariants. \cite{graf2013bulk,graf2018bulk}.

\end{rk}

\paragraph{Sketch of the proof}
Let us denote by $A$ the operator $A=C\theta(X)(1-S^2)$ appearing in the edge index expression. Consider the anti-commutator $\{A,S\} = C[\theta(X),S](1-S^2)$. The proof of the main theorem relies, on the one hand, on the fact that $[\theta(X),S]$ is exponentially small when $x$ or $y$ is far from the switch of $\theta$ and on the other hand that $(1-S^2)_{x,y}$ is exponentially small when $x$ or $y$ is far from the edge. Thus if the switch of $\theta$ is far enough from the edge, then $\{A,S\}$ has exponentially small matrix elements, see Proposition~\ref{prop:anticommutator} below. 
We see also that $A-A^3=C\theta(X)\left((1-S^2)-(1-S^2)^3\right) + B$ where $B$ is an operator involving some commutator of $1-S^2$ and $\theta$. Therefore $B$ is also exponentially small for the same reason. 

If we allow ourselves to neglect those exponentially small terms, we would obtain that $\{A,S\}=0$ and $S\ket{\psi}=0 \Rightarrow (A-A^3)\ket{\psi}=0$ (which implies $(A-A^3)\ket{\psi}\neq0 \Rightarrow S\ket{\psi}\neq 0$). Therefore, to each eigenstate $\ket{\psi}$ of $A$ with eigenvalue $\lambda$ not in $\lambda\in \{0,1,-1\}$, we could associate an eigenvector $S\ket{\psi}$ of $A$ with opposite eigenvalue ($AS\ket{\psi}=-SA\ket{\psi}= -\lambda S\ket{\psi}$) and same multiplicity. So the contribution of all eigenvalues not in $\{0,1,-1\}$  will cancel out two by two and the trace of $A$ would read $\dim \ker (A-1) - \dim \ker (A+1) \in \mathbb Z$. 

The additional difficulty  with the complete proof of Theorem~\ref{thm:main} is to keep track and bound rigorously those error terms and show that they induce only a small deviation to the quantization of the edge index.

\subsection{A numerical example}

We illustrate our results on a non-homogeneous version of the SSH chain given by the equation~\eqref{eq:SSH}, by considering for $x\in [1,L]$
$$
t_{1,x}= \frac{1}{2} + \tilde{t}^{\text{rd}}_{1,x} + t^\mathrm{defect}_x,\qquad t_{2,x} = 1 + \tilde{t}^{\text{rd}}_{2,x}
$$
where $\tilde{t}^{\text{rd}}_{1,x}$ and $\tilde{t}^{\text{rd}}_{2,x}$ are single disordered configurations from independent random variables, identically distributed with a uniform law supported in $[-0.1,0.1]$ and $t^\mathrm{defect}_x = 0.2 \exp(-(4x/L)^2)$ is a Gaussian-shape defect in the middle of the chain. 

\begin{figure}[htb]
    \centering
    \includegraphics[height=4cm,width=8cm]{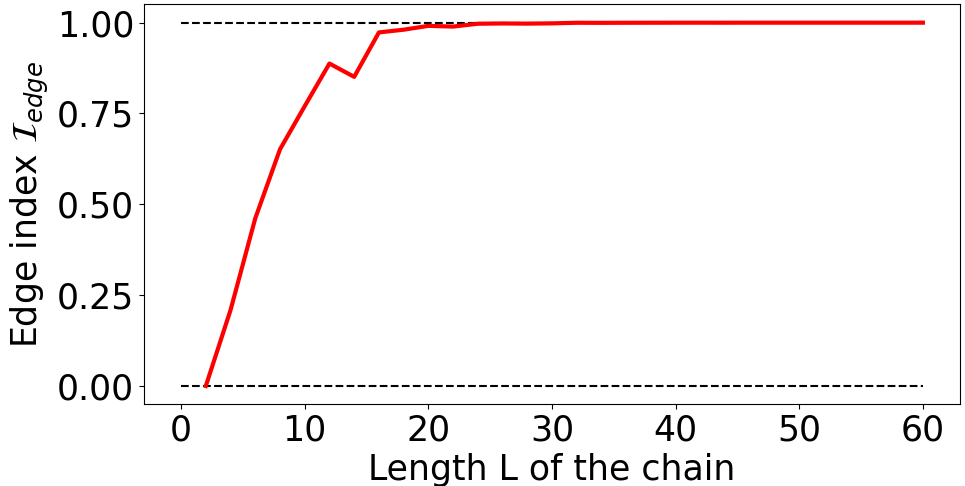}
    \includegraphics[height=4cm,width=8cm]{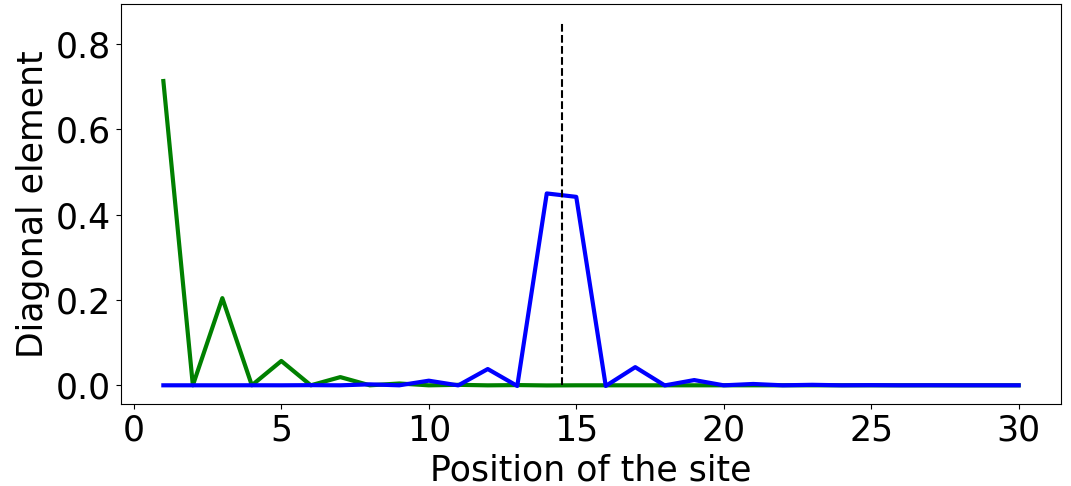}
    \caption{Left: Numerical value of $\mathcal I_\mathrm{edge}$ with respect to the system size $L$ ($\delta=1/\sqrt{2L}$). Right: Diagonal elements of the matrices $CS[\theta(X),S]$ (in blue) and $C\theta(X)(1-S^2)$ (in green) appearing in bulk and edge index expressions of Definition~\ref{def:indices} ($L=30,\,\delta=1/20$).}
    \label{fig:numerical_edge}
\end{figure}
The numerical value of the edge index, computed according to Definition~\ref{def:indices}, is plotted with respect to $L$ in the left panel Figure~\ref{fig:numerical_edge}. We see a fast exponential convergence towards $\mathcal I_\mathrm{edge}=1$ as predicted by Theorem~\ref{thm:main}. In the right panel, we plot the 
value of the diagonal elements of the matrices $CS[\theta(X),S]$ and $C\theta(X)(1-S^2)$, whose respective trace gives the bulk and edge index. As expected, we observe that the non-vanishing contributions to these quantities are localized in space, respectively near the transition of $\theta$ and near the edge (as sketched in Figure~\ref{Schema}).

Finally, we also study the influence on the edge index of the position of the jump of $\theta$ and the choice of $\delta$. The former does not matter as long as it is far away enough from the edges. For the latter, we see that $\delta$ must be in a certain interval for $\mathcal I_\mathrm{edge}$ to be close enough to an integer, in agreement with the qualitative discussion of Remark~\ref{rk:delta}.

\section{Proof of the main Theorem}

\subsection{Proof of Proposition~\ref{prop:BEC} \label{sec:proofBEC}}

The proof of this result is very elementary. It suffices to use the anti-commutation relation with the chirality operator $CS = C\tanh(H/\delta)=\tanh(-H/\delta)C=-SC$ as well as the cyclicity of the trace to rearrange the terms in the following order:
\begin{equation}
	\begin{aligned}
		\mathcal{I}_{\mathrm{edge}}&=\Tr(C\theta(X)(1-S^2)) = \Tr(C\theta(X)) - \Tr(C\theta(X)S^2)\\
		&=\Tr(C\theta(X)) + \frac{1}{2}\Tr(C[S,\theta(X)]S)=\Tr(C\theta(X)) +\mathcal{I}_{\mathrm{bulk}}.
	\end{aligned}
\end{equation}
Finally, $\Tr(C\theta(X))=0$ since $\theta(X)$ acts trivially on $\mathbb C^2$, which proves the proposition.

In the case where sublattice sites are encoded in the parity of the lattice position, one has instead $\Tr(C\theta(X)) = n_{A,\theta}-n_{B,\theta}$ as $C$ takes value $+1$ on the $A$ sites and $-1$ on the B sites, in agreement with Remark~\ref{rk:SSH}.

\subsection{Preliminary results}

We establish several auxiliary results that will be used in the proof of Theorem~\ref{thm:main}. They are also of independent interest and justify the ``bulk'' and ``edge'' terminology for the indices.

\paragraph{Inequality notation.} In the rest of the paper, most of the inequalities will be governed by exponential decay. Therefore we will want to simply the computations by using the notation  $$g=\mathcal{O}\left(h\right)$$ when there is a function $\gamma$ such that $|g|\leq \gamma h$ with $\gamma$ a polynomial function in the variables $(K_d,d,\delta,1/\delta,L)$. 


\paragraph{Weyl functional calculus.}

The goal of functional calculus is to define what is $f(H)$ where $f$ is function on the spectrum of some operator $H$. These operators appear often in physics like the evolution operator $e^{iHt}$ or the thermal density of states $P= \frac{1}{1+e^{H/(k_bT)}}$. For matrices, the natural way to is to define $f(H) = \sum_i f(\lambda_i) \ket{\psi_i}\bra{\psi_i}$ when the eigen-decomposition of $H$ reads $H=\sum_i \lambda_i \ket{\psi_i}\bra{\psi_i}$. Here instead we shall use the Weyl formulation \cite{anderson1969weyl}:

\begin{equation}
    f(H) = \frac{1}{2\pi}\int_{-\infty}^{\infty} d\omega \hat{f}(\omega) e^{i\omega H}
\end{equation}
where $\hat{f}$ is the Fourier transform of $f$ and the evolution operator $e^{i\omega H}$ is defined using the usual equation $\partial_\omega e^{i\omega H}= iHe^{i\omega H}$. This definition coincides with the previous one when $f$ is smooth.

With this formulation, we prove that when $f$ is a smooth function and $H$ a short range operator, then $f(H)$ is also short range.



\begin{prop}\label{prop:combes-thomas}
Let $H$ be a Hamiltonian that satisfies Assumption~\ref{hyp:shortrange}. Suppose that $f$ is regular enough such that there exists  $\beta>0$ such that $\sup_{\omega \in \mathds{R}}  |\hat{f}(\omega) \omega| < C_\beta e^{-\beta |\omega|}$. Then
\begin{equation}
    \norm{f(H)_{x,y}}  \leq 4\left(\|f\|_\infty+\frac{C_\beta K_d}{\beta}+C_\beta\frac{|x-y|}{d}\right) e^{-|x-y|/d'}
\end{equation} for $d'=d\max(1,K_d/\beta)$ and $\|f\|_\infty=\sup\{f(x),x\in \mathds{R}\}$.
\end{prop}

The proof can be found in Section~\ref{sec:proof-combes-thomas} below. It relies on Lieb-Robinson bound \cite{hastings2010locality}, which says that  $e^{i\omega H}$ is short range and on some Fourier analysis of $f$. This result is nothing but a consequence of Combes-Thomas estimate. 


Now If we take $f(z) = \tanh(z/\delta)$, whose Fourier transform is $\hat{f}(\omega)= -i\delta\sqrt{\pi/2}\csch(\pi\delta\omega/2)$, we can check that for $\beta = \pi \delta/4$ we have the requested property for some finite $C_\beta=1.5$. Therefore we deduce
\begin{cor}
Let $H$ be a Hamiltonian that satisfies Assumption~\ref{hyp:shortrange} and  $S=(\tanh(H/\delta))$ for $\delta>0$. We have
\begin{equation}
    \norm{S_{x,y}}
    =\mathcal{O}( e^{-|x-y|/(2d')})
\end{equation}
 with $d' = d\max(1,4 K_d/(\pi \delta))$.
 \end{cor}

\begin{rk}[Localization of bulk expression]
If we consider the commutator $[\theta,S]$ then $[\theta,S]_{x,y}=S_{x,y} (\theta(y)-\theta(x)) $ is only non-negligible when $x,y$ is around the switch of $\theta$. Thus in the bulk index expression
\begin{equation}
	\mathcal{I}_{\text{bulk}} = \frac{1}{2}\Tr\left(CS[\theta(X),S]\right) = \sum_{x,y} \frac{1}{2} C(x) S_{y,x}S_{x,y}(\theta(y)-\theta(x))
\end{equation}
the only terms $S_{x,y}$ that contribute significantly to the trace are those close to the transition of $\theta$, as illustrated in Figure~\ref{Schema}. This justifies that this index is of "bulk" type.
\end{rk}

Another important property is that, away from de edges of the chain, any smooth enough function of $H$ is close to the bulk Hamiltonian $H^{\text{bulk}}$ \cite{loring2021locality}. Recall that $\iota : \HH_L \to \HH^\mathrm{bulk}$ is the canonical inclusion and $\iota^* : \HH^\mathrm{bulk} \to \HH_L$ the canonical restriction, so that $H=\iota^* H^\mathrm{bulk} \iota$.
 
\begin{prop}\label{prop:edge-loc}
Let $f$ be a regular function such that there exist a $\beta>0$ verifying $\sup_{\omega \in \mathds{R}}  |\hat{f}(\omega) \omega| < C_\beta e^{-\beta |\omega|}$. Let $\Omega \subset \llbracket 0, L-1 \rrbracket$ be a sub-region of the chain and let $\chi_\Omega$ be the characteristic function associated to $\Omega$. If we denote by $d_\Omega$ the distance between $\Omega$ and the edges $\{0,L-1\}$ of the chain, we have
 \begin{equation}
     \|\chi_\Omega (f(H)- f(\iota^* H^{\mathrm{bulk}}\iota) )\| \leq 4C_\beta\left(\frac{ d_\Omega N^2_d}{d}+1+\|f\|_\infty\right)e^{-d_\Omega/(2d')}
 \end{equation}
with  $d'=d\max(1,K_d/\beta)$ and $N_d =\sup_x \sum_{y}e^{-|x-y|/(2d)}$
 \end{prop}
Notice that the left-hand side difference can be written $f(\iota^* H^{\text{bulk}} \iota)-\iota^* f(H^{\text{bulk}}) \iota$, so that this proposition actually compares how functional calculus and restriction to the open chain do not commute. Because the operators are short-range, this difference is exponentially small away from the boundary. 

Taking $f(E) =1- \tanh^2(E/\delta)$, $f$ satisfies the regularity hypothesis for $\beta =\pi \delta/4$ and  $C_\beta=3$. Since $H^\mathrm {bulk}$ has a spectral gap (see Assumption~\ref{hyp:gap}) we have:
$$
\norm{ f(H^{\text{bulk}})} = \sup\left\{|1- \tanh^2(E/\delta)|,E \in \Sp(H^{\text{bulk}})\right\}\leq \sup\left\{|1- \tanh^2(E/\delta)|,|E| \geq \Delta \right\} \leq 4e^{-2\Delta/\delta}
$$
In particular for $x \in \mathbb \llbracket 0,L-1 \rrbracket$ we consider $\Omega=\{x\}$ and denote (with a slight abuse) $d_x = \min(x,L-1-x)$ the distance between $x$ and the edges.  
 
 \begin{cor} If $H^{\text{bulk}}$ satisfies Assumption~\ref{hyp:gap} then, for $\delta>0$, $S=(\tanh(H/\delta))$ satisfies:
\begin{equation}
    \norm{(1-S^2)_{x,y}} =\mathcal{O} \left(e^{-\max(d_x,d_y)/(2d')} +e^{-2\Delta/\delta}\right)\label{edge decay}
\end{equation}
where $d'=d\max(1,4K_d/(\pi\delta))$ and $d_x$ denote the distance of a site $x$ to the edges of the chain.
\end{cor}

\begin{rk}[Localization of edge expression]The windowed density of states $(1-S^2)_{x,x}$ quickly decays for $x$ far from the edges. Therefore when we compute the edge index:
\begin{equation}
    \mathcal{I}_{\text{edge}}= \Tr(C\theta(1-S^2)) = \sum_x C(x)\theta(x)(1-S^2)_{x,x}
\end{equation}
the only terms in the sum that will be relevant are those which are close to the edge, as illustrated in Figure~\ref{Schema}. This justify calling this index of "edge" type. Moreover the switch function $\theta$ ensures that we compute the contribution of the left edge only.

Proposition~\ref{prop:edge-loc} also guarantees that for large enough chains ($L\gg d'$), computing the edge index for finite chains or semi-infinite chains give the same result (up to exponentially small deviations). Therefore it ensures that the finite edge index will be close to the integer value of its infinite counter-part (excluding the pathological behavior where the finite index is zero even if the semi-infinite system is topological).

\end{rk}

The last result combines the previous ones and is central in the proof of the main theorem. We recall that for a matrix $A$ of size $n$ the trace norm is defined by:
$$
\norm{A}_1 = \Tr(\sqrt{A^*A}) = \sum_{i=1}^n \sigma_i(A),
$$
where $\sigma_i(A)$ are the singular values of $A$ and satisfies $|\Tr(A)| \leq \norm{A}_1$.

\begin{prop}\label{prop:anticommutator}
	Consider $A=\frac{1}{2}C\{\theta(X),(1-S^2)\}$. Then $\{A,S\} = C\{[\theta(X),S],(1-S^2)\}$ satisfies:
	\begin{equation}
		||\{A,S\}||_1=\mathcal{O}\left(e^{-2\Delta/\delta}+e^{-L/(48d')}\right)
	\end{equation}
where $d'=d\max(1,4K_d/(\pi\delta))$. Similarly, one has:
$
\norm{[1-S^2,\theta]}_1=\mathcal{O}\left(e^{-2\Delta/\delta}+e^{-L/(48d')}\right).
$
\end{prop}

\subsection{Proof of Theorem~\ref{thm:main}}

In order to show that $\mathcal{I}_\text{edge}$ is almost an integer we will  show that 
$$
e^{i2\pi\mathcal{I}_\text{edge}}= e^{i2\pi \Tr(A)}=\det(e^{i2\pi A}), 
$$
is almost equal to one. For that we want to use the anti-commutation between $A$ and $S$ and so we artificially introduce the product $\frac{S+i\epsilon}{S+i\epsilon}$. The parameter $\epsilon$ is here to regularize the expression (as $S^{-1}$ is not defined in general) and will be carefully chosen later.

\begin{equation}
	\begin{aligned}
		\det(e^{i2\pi A})&=\det(\frac{1}{S+i\epsilon})\det(e^{i\pi A})\det(S+i\epsilon)\det(e^{i\pi A})=\det(\frac{1}{S+i\epsilon}e^{i\pi A}(S+i\epsilon)e^{i\pi A})\\
		&=\det(\frac{1}{S+i\epsilon}(e^{i\pi A}S-Se^{-i\pi A})e^{i\pi A}+ \frac{i\epsilon}{S+i\epsilon}e^{i2\pi A}+\frac{S}{S+i\epsilon})\\
		&=\det(1+\frac{1}{S+i\epsilon}(e^{i\pi A}S-Se^{-i\pi A})e^{i\pi A}+ \frac{i\epsilon}{S+i\epsilon}(e^{i2\pi A}-1))
	\end{aligned}\label{19}
\end{equation}

Now we want to prove that the terms which are not the identity are small and thus that the determinant only slightly deviate from 1. In order to do that we will use the following lemma:

\begin{Lemma}
	If $T$ is an operator such that $||T||_1 <1$ then:
	\begin{equation}
		|\det(1+T)-1|\leq \frac{||T||_1}{1-||T||_1}
	\end{equation}
\end{Lemma}
\begin{Proof}
	We have that $\det(1+T)-1 = \int_0^1dt \partial_t \det(1+tT)=\int_0^1dt\Tr(T(1+tT)^{-1})$ which lead to the following inequality $|\det(1+T)-1| \leq \frac{||T||_1}{1-||T||}\leq \frac{||T||_1}{1-||T||_1}$
\end{Proof}

So we want to prove that the norms $||\cdot||_1$ of the right two terms of \eqref{19} are small. For the term $\frac{1}{S+i\epsilon}(e^{i\pi A}S-Se^{-i\pi A})e^{i\pi A}$ it can be down relatively easily once we know that $||\{S,A\}||_1$ is small by Proposition~\ref{prop:anticommutator}:
\begin{align}
	||\frac{1}{S+i\epsilon}(e^{i\pi A}S-Se^{-i\pi A})e^{i\pi A}||_1 &\leq \frac{1}{\epsilon}||e^{i\pi A}Se^{i\pi A}-S||_1 \cr & \leq \frac{1}{\epsilon}||\int_0^{1}dt \pi e^{i \pi t}\{S,A\}e^{i\pi t}||_1 \cr & \leq \frac{\pi}{\epsilon} ||\{S,A\}||_1
\end{align}

For the second term $\frac{i\epsilon}{S+i\epsilon}(e^{i2\pi A}-1)$ we need a little bit more work. First if we denote by $B_t= \cos(2\pi t(1-S^2))\theta+i\sin(2\pi t(1-S^2))C\theta +1-\theta$ we will show that $||B_1-e^{i2\pi A}||_1$ is small for that we show that:
\begin{align}
		\partial_t(e^{i2\pi tA}-B_t) &= e^{it2\pi A}i2\pi A-\sin(2\pi(1-S^2))2\pi(1-S^2)\theta+\cos(2\pi(1-S^2))i2\pi(1-S^2)C\theta\cr
		&=\left(e^{i2\pi tA}-B_t\right)i2\pi A +\left(\cos(2\pi(1-S^2))i C-\sin(2\pi(1-S^2))\right)2\pi [1-S^2,\theta]
\end{align}
which implies:
\begin{align}
		\|e^{i2\pi tA}-B_1\| &= \| \int_0^1dt e^{-i2\pi tA}2\pi \left(\cos(2\pi(1-S^2))i C-\sin(2\pi(1-S^2)) \right)[1-S^2,\theta]\| \cr
		&\leq 4\pi ||[1-S^2,\theta]||_1
\end{align}

 Moreover, by Proposition~\ref{prop:anticommutator} we also have $||[1-S^2,\theta]||_1=\mathcal{O}\left(e^{-2\Delta/\delta}+e^{-L/(24d')}\right)$. So if we decompose $e^{i2\pi A}-1$ as $(e^{i2\pi A}-B_1)+(B_1-1)$ we obtain:
	\begin{align}
		||\frac{i\epsilon}{S+i\epsilon}(e^{i2\pi A}-1))||_1 &\leq 4\pi ||[1-S^2,\theta]||_1+ ||\frac{i\epsilon}{S+i\epsilon}(B_1-1))||_1\cr
		&\leq 4\pi ||[1-S^2,\theta]||_1+ \epsilon\left(||\frac{\cos(2\pi(1-S^2))-1}{S+i\epsilon}\theta||_1 +||\frac{\sin(2\pi(1-S^2))}{S+i\epsilon}\theta||_1\right)\cr
		&\leq 4\pi ||[1-S^2,\theta]||_1+ \epsilon\left(||\frac{\cos(2\pi S^2)-1}{S+i\epsilon}\theta||_1+||\frac{\sin(2\pi S^2)}{S+i\epsilon}\theta||_1\right),
	\end{align}
where $\cos(2\pi(1-S^2))=\cos(2\pi S^2)$ come from the usual properties of the $\cos$ applied to all the eigenvalues. If we then introduce the function $f(x) = \frac{\cos(2\pi x^2)-1}{x}$ and $g(x)= \frac{ \sin(2\pi x^2)}{x}$ we see that the right term can be re-express as:
\begin{equation}
	\begin{aligned}
||\frac{\cos(2\pi S^2)-1}{S+i\epsilon}\theta||_1+||\frac{\sin(2\pi S^2)}{S+i\epsilon}\theta||_1 &\leq ||\frac{S}{S+i\epsilon} f(S)\theta||_1+||\frac{S}{S+i\epsilon} g(S)\theta||_1\\
		&\leq (||f(x)||_\infty+||g(S)||_\infty)\|\theta\|_1= \mathcal{O}\left(1\right)\\
	\end{aligned}
\end{equation}

Therefore at the end we have that:
\begin{equation}
	|e^{i2\pi \mathcal{I}_\text{edge}}-1| = (\frac{1}{\epsilon}+1)\mathcal{O}\left(e^{-2\Delta/\delta}+e^{-L/(48d')}\right)+\epsilon \mathcal{O}\left(1\right)
\end{equation}

If we take $\epsilon = \sqrt{e^{-2\Delta/\delta}+e^{-L/(48d')}}$  we thus obtain that:
\begin{equation}
	|e^{i2\pi \mathcal{I}_\text{edge}}-1| = \mathcal{O}\left(e^{-\Delta/\delta}+e^{-L/(96d')}\right)
\end{equation}
which gives us the result:
\begin{equation}
	\mathcal{I}_\text{edge} \in \mathds{Z} + \mathcal{O}\left(e^{-\Delta/\delta}+e^{-L/(96d')}\right)
\end{equation}
with $d' =d\max(1,4K_d/(\pi\delta))$.

Since $96\times \frac{4}{\pi}\leq 128$ and choosing $\delta= \sqrt{\frac{\Delta dK_d128}{ L}}$ we finally find the claimed result:
\begin{equation}
	\mathcal{I}_\text{edge} \in \mathds{Z} + \mathcal{O}\left(e^{-\sqrt{\frac{ L\Delta}{128K_d d}}}\right)
\end{equation}

\section{Remaining proofs}

\subsection{Proof of Proposition~\ref{prop:combes-thomas} \label{sec:proof-combes-thomas}}

We proceed in two steps. First we will prove a Lieb-Robinson-like bound which is adapted to the majoration we need for our problem. Then we will use Weyl functional calculus to extend the exponential decay property of the operator $e^{itH}$ to a wider class of functional operator $f(H)$.

To begin let $z$ be some arbitrary point of the lattice. Let then denote by $M_z$ the diagonal operator that act on the basis of sites as $M_z \ket{x} = e^{|x-z|/d} \ket{x}$. Then if we denote by $B$ the operator $B=M_zHM^{-1}_z-H$ we see that:
\begin{equation}
	|B_{x,y}| \leq |H_{x,y}| (e^{ (|y-z|-|x-z|)/d'}-1)\leq |H_{x,y}| (e^{ (|y-x+x-z|-|x-z|)/d}-1)\leq  |H_{x,y}| (e^{|y-x|/d}-1)
\end{equation}
Then we will use the general fact that $||B|| \leq \sqrt{\sup_x \sum_y |B(x,y)|\sup_y \sum_x |B(x,y)|}$ to obtain that $||B||\leq K_{d}$. This fact can be obtained by showing that if there is ($\lambda,\psi$) such that $\lambda \psi = B^\dagger B \psi$ then we have:
\begin{equation}
	|\lambda|\sup_x| \psi_x|=\sup_x|(B^\dagger B \psi)_x|\leq \sup_x \sum_y |B(x,y)|\sup_y \sum_x |B(x,y)| \sup_x|\psi_x|
\end{equation}
which implies that $|\lambda|\leq \sup_x \sum_y |B(x,y)|\sup_y \sum_x |B(x,y)|$ and therefore that:
\begin{equation}
	\|B\|= \|\sqrt{B^\dagger B}\| =\sqrt{\sup \{|\lambda|, \lambda \in \Spec(B^\dagger B)\}}\leq \sqrt{\sup_x \sum_y |B(x,y)|\sup_y \sum_x |B(x,y)|} 
\end{equation}


Now that we have that $\|B\| \leq K_d$ we can use that $\partial_t(e^{it(H+B)}e^{-itH})=e^{it(H+B)}Be^{-itH}$ to show that:
\begin{equation}
	\begin{aligned}
		e^{it(H+B)}-e^{itH}&= \int_0^tds \left(e^{is(H+B)}-e^{isH}+e^{isH}\right)Be^{i(t-s)H}\\
		\Rightarrow \|e^{it(H+B)} -e^{itH}\| &\leq \int_0^t ds \|e^{is(H+B)}-e^{isH}\| \|B\| + \|B\|t
	\end{aligned}\label{20}
\end{equation}
which by Gr\"onwall's inequality implies that $\|e^{it(H+B)} -e^{itH}\|\leq |t|\|B\|e^{|t|\|B\|}\leq |t|K_de^{|t|K_d}$. Then we use that $e^{it(H+B)}-e^{itH} =e^{itM_zHM_z^{-1}} -e^{itH} = M_ze^{itH}M_z^{-1} -e^{itH}$ and if we then take $z=y$ we see that $(M_ye^{itH}M_y^{-1}-e^{itH})_{x,y} = (e^{|x-y|/d}-1)(e^{itH})_{x,y}$ and therefore the previous inequality imply that $|(e^{itH})_{x,y}| \leq |t| K_d e^{|t|K_d}/\left(e^{|x-y|/d}-1\right)$. If we only look for large distance where $|x-y| \geq d$, it reduces to: 
\begin{equation}\label{LiebRobinson}
	|(e^{itH})_{x,y}| \leq 2|t|K_d e^{|t|K_d-|x_y|/d}
\end{equation} 
which is an inequality of the Lieb-Robinson type.

Now we want to study the operator $f(H) = \int_{-\infty}^\infty d\omega \hat{f}(\omega) e^{i\omega H}$ and show that it coefficients $f(H)_{x,y}$ decay exponentially fast for long distance $|x-y|\gg d$. For that we will introduce an arbitrary parameter $\alpha>0$ and use different inequalities for majoring  depending on if we work with small $|\omega|<\alpha$ or large ones $|\omega|\geq\alpha$. For the small one we will use \eqref{LiebRobinson}
and for the big ones we will use $|(e^{i\omega H})_{x,y}|\leq \|e^{i\omega H}\|=1$. On the other hand we will use the supposed majoration for $\hat{f}(\omega)$ which is $|\hat{f}(\omega)| \leq C_\beta e^{-\beta \omega}/|\omega|$. All this together gives us the following inequalities for $|x-y|\geq d$:
\begin{equation}
	\begin{aligned}
		f(H)_{x,y} &= \int_{-\infty}^\infty d\omega \hat{f}(\omega) (e^{i\omega H})_{x,y}\\
		| f(H)_{x,y}| & \leq \int_{-\infty}^\infty d\omega\mathds{1}_{\{|\omega| \leq \alpha\}}|\hat{f}(\omega)|2|\omega|K_d e^{|\omega|K_d-|x-y|/d}+ \int_{-\infty}^\infty d\omega\mathds{1}_{\{|\omega| \geq \alpha\}} |\hat{f}(\omega)|\\
		&\leq \int_{-\infty}^\infty d\omega\mathds{1}_{\{|\omega| \leq \alpha\}}|2C_\beta K_d e^{|\omega|(K_d-\beta)-|x-y|/d}+ \int_{-\infty}^\infty d\omega\mathds{1}_{\{|\omega| \geq \alpha\}} \frac{C_\beta}{\alpha}e^{-\beta |\omega|}\\
		&\leq 4C_\beta K_d \alpha \max(1,e^{(K_d-\beta)\alpha})e^{-|x-y|/d} + \frac{2C_\beta}{\alpha \beta} e^{-\beta \alpha}
	\end{aligned}
\end{equation}

To obtains one of the tighter inequalities we choose $\alpha  =|x-y|/(dK_d)$ and we therefore obtains that for $|x-y|\geq d$:

\begin{equation}
	\begin{aligned}
		| f(H)_{x,y}| &\leq \frac{4C_\beta|x-y|}{d}e^{-\min(1,\beta/K_d) |x-y|/d}+2C_\beta K_d/\beta e^{-\beta/K_d|x-y|/d} \\*
		&\leq 4C_\beta \left(K_d/\beta+|x-y|/d\right)e^{-\min(1,\beta/K_d)|x-y|/d}   
	\end{aligned}
\end{equation}
This inequality is a valid only for $|x-y|\geq d$, but for the small distance we can just use the fact that $f(H)_{x,y}\leq\|f(H)\| \leq \|f\|_\infty
$ to deduce that uniformly in $x,y$ we have the inequality:

\begin{equation}
	| f(H)_{x,y}| \leq 4\left((\|f\|_\infty+C_\beta K_d/\beta+C_\beta|x-y|/d)\right) e^{-\min(1,\beta/K_d)|x-y|/d}   
\end{equation}
which ends the proof.

\subsection{Proof of Proposition~\ref{prop:edge-loc}}

In all this section, to simplify the notations we will denote by  $\Tilde{H}$ the operator $\iota H\iota^*$.

Consider again Equation~\eqref{20} from the  proof of Proposition~\ref{prop:combes-thomas} and bound it instead in the following way:
\begin{equation}
	\begin{aligned}
		e^{it(H+B)}&= e^{itH} + \int_0^tds e^{is(H+B)}Be^{i(t-s)H}\\
		\Rightarrow \|e^{it(H+B)} \| &\leq 1+\|B\| \int_0^t ds  \|e^{is(H+B)} \|
	\end{aligned}
\end{equation}
which by Gr\"onwall inequality gives that $\|e^{it(H+B)} \| \leq e^{|t|\|B\|}$ and as $\left(e^{it(H+B)}\right)_{x,y}=\left(M_z e^{itH}M_z^{-1}\right)_{x,y}= (e^{itH})_{x,y}e^{|x-y|/d}$ for $z=y$ we recover another Lieb-Robinson inequality, this time valid for every $x,y$:
\begin{equation}
	|(e^{itH})_{x,y}| \leq e^{|t|K_d-|x-y|/d}
\end{equation}

And then we use this inequality to derive that:
\begin{equation}
	\begin{aligned}
		\|\chi_\Omega(e^{itH^{\text{bulk}}}-e^{it\Tilde{H}})\| &= \|\chi_\Omega\int_0^s ds e^{isH^{\text{bulk}}} (H^{\text{bulk}}-\Tilde{H})e^{i(t-s)\Tilde{H}}\|\leq \int_0^s\| \chi_\Omega e^{isH^{\text{bulk}}} (H^{\text{bulk}}-\Tilde{H})\|\\
		&\leq \int_0^s ds\sup_x \sum_{y,z} | \chi_{\Omega,z} (e^{isH^{\text{bulk}}})_{y,z} (\Tilde{H}_{x,y} -H^{\text{bulk}}_{x,y})| \\
		&\leq |t| \sup_x \sum_{y,z} e^{|t|K_d-|z-y|/d}2K_de^{-|x-y|/d}\chi_{\Omega,z}\mathds{1}_{\Tilde{H}_{x,y}\neq H_{B,x,y}}
	\end{aligned}
\end{equation}

where we used $|(e^{itH^{\text{bulk}}})_{x,y}| \leq e^{|t|K_d-|x-y|/d}$ and $|H_{x,y}| \leq K_de^{-|x-y|/d}$ (which is a direct consequence of Assumption \ref{hyp:shortrange}). Then if we use that $\chi_{\Omega,z} \mathds{1}_{\Tilde{H}_{x,y}\neq H^{\text{bulk}}_{x,y}}\neq 0$ only if $x$ or $y$ is out of the support of $\theta$ and $z$ in the support of $\chi_\Omega$, we can deduce that $|x-y|+|y-z|\geq d_\Omega$ leading to:
\begin{equation}
	\begin{aligned}
		\|\chi_\Omega(e^{itH^{\text{bulk}}}-e^{it\Tilde{H}})\| &\leq |t|(2K_d+\|\Tilde{H}\|+\|H^{\text{bulk}}\|) e^{|t|K_d-d_\Omega/(2d)}\sup_x \sum_{y,z}e^{-(|x-y|+|y-z|)/(2d)}\\
		&\leq |t|2K_d e^{|t|K_d-d_\Omega/(2d)} N_d^2
	\end{aligned}
\end{equation}
where $N_d= \sup_x \sum_{y}e^{-|x-y|/(2d)}$. Once we have prove this inequality for the propagator, we use the Weyl calculus to extend it to the smooth function case:

\begin{equation}
	\begin{aligned}
		&\|\chi_\Omega(f(H^{\text{bulk}})-f(\Tilde{H}))\|  \leq \int_{-\infty}^\infty d\omega |\hat{f}(\omega)|\|\chi_\Omega(e^{iH^{\text{bulk}}}-e^{i\omega \Tilde{H}})\|\\
		&\leq \int_{-\infty}^\infty d\omega |\hat{f}(\omega)|\mathds{1}_{\{|\omega| \leq \alpha\}}|\omega|2K_d e^{|\omega|K_d-d_{\theta\theta'}/(2d)} N_d^2+\int_{-\infty}^\infty d\omega 2|\hat{f}(\omega)|\mathds{1}_{\{|\omega| \geq \alpha\}}\\
		&\leq 4C_\beta\alpha2K_dN^2_d\max(1,e^{(K_d-\beta)\alpha})e^{-d_{\theta\theta'}/(2d)}+\frac{ 4C_\beta}{\beta \alpha} e^{-\beta\alpha}
	\end{aligned}
\end{equation}
then we take $\alpha =d_\Omega/(2dK_d) $ to obtain that:
\begin{equation}
	\begin{aligned}
		\|\chi_\Omega(f(H^{\text{bulk}})-f(\Tilde{H}))\|  &\leq C_\beta\left(\frac{4 d_\Omega N^2_d}{d}+\frac{ 8dK_d}{\beta d_\Omega}\right)e^{-\min(1,\beta/K_d)d_\Omega/(2d)}
	\end{aligned}
\end{equation}

One can remove the singularity in $1/d_\Omega$ by also using the inequality $\|\chi_\Omega(f(H_B)-f(H))\|\leq \|f\|_\infty$ for $d_\Omega\leq 2d\max(1,\beta/K_d)$ leading to the inequality:

\begin{equation}
	\begin{aligned}
		\|\chi_\Omega(f(H^{\text{bulk}})-f(\Tilde{H}))\|  &\leq C_\beta\left(\frac{4 d_\Omega N^2_d}{d}+4\min(1,K_d/\beta)+e\|f\|_\infty \right)e^{-\min(1,\beta/K_d)d_\Omega/(2d)}\\
		&\leq 4C_\beta\left(\frac{ d_\Omega N^2_d}{d}+1+\|f\|_\infty\right)e^{-\min(1,\beta/K_d)d_\Omega/(2d)}
	\end{aligned}
\end{equation}

To arrive to the  inequality from in the Proposition we  use that the projecting on the finite chain using $\iota$ is norm decreasing meaning that $\|\iota^* \chi_\Omega(f(H^{\text{bulk}})-f(\Tilde{H}))\iota\|\leq \|\chi_\Omega( f(H^\mathrm{bulk})-f(\Tilde{H}))\|$. Then we use that because $\iota^* \iota =1$ we have that $\iota^*f(\Tilde{H})\iota=\iota^*f(\iota H \iota^*)\iota= f(H)$ (this property can be checked manually for polynomials or exponentials and extended by density to all continuous functions).

\subsection{Proof of Proposition~\ref{prop:anticommutator}}

We shall use of the following lemma whose proof can be found in \cite[Lemma 11]{graf_bulk-edge_2018}.
\begin{Lemma}
	Let $T$ be an operator acting on an Hilbert space $\mathcal{H}$ and let $(\ket{x})_x$ be an orthonormal basis of $\mathcal{H}$. Then if we denote by $||T||_1$ the norm $\Tr(|T|)$ and by $T_{x,y}$ the coefficient $\bra{y}T\ket{x}$, then we have the inequality:
	\begin{equation}
		||T||_1\leq \sum_{x,y} |T_{x,y}|
	\end{equation}
\end{Lemma}
Using this lemma with $\{A,S\}_{x,y} = C\{[\theta(X),S],(1-S^2)\}_{x,y}$  we obtain:
\begin{equation}
	\begin{aligned}
		||\{A,S\}||_1 &\leq \sum_{x,y} \left(C\left\{[\theta(X),S],(1-S^2)\right\}\right)_{x,y} \leq \sum_{x,y,z} 2 \left|[\theta(X),S]_{z,y}(1-S^2)_{x,z}\right|\\
		&\leq \sum_{x,y,z} \mathcal{O}\left( e^{-(|z-y|/(2d')}\mathds{1}_{\theta_z \neq \theta_y} \min\left(e^{-\max(d_x,d_y)/(2d')} +e^{-2\Delta/\delta},e^{-|x-y|/(2d')}\right)\right)\\
	\end{aligned}
\end{equation}
Where $\mathds{1}_{\theta_z \neq \theta_y}$ is the characteristic function in $x,y$ associated to the condition $\theta_x\neq\theta_y$.
We will then introduce a free parameter $d_\lambda$ such that when $|x-y|\leq d_\lambda$ we use of $|(1-S^2)_{x,y}|$the bound by $e^{-\max(d_x,d_y)/(2d')} +e^{-2\Delta/\delta}$ and when $|x-y|\geq d_\lambda$ we use the one by $e^{-|x-y|/(2d')}$. This then gives:
\begin{align}
		||\{A,S\}||_1 &\leq \mathcal{O}\Big(\hspace{-0.3cm}\sum_{\substack{x,y,z\\ {\scriptscriptstyle|x-y|\leq d_\lambda}}} \hspace{-0.2cm} e^{-(|z-y|/(2d')}\mathds{1}_{\scriptscriptstyle\theta_z \neq \theta_y} \left(e^{-\max(d_x,d_y)/(2d')}
		+e^{-2\Delta/\delta}\right)+\hspace{-0.3cm}\sum_{\substack{x,y,z\\ {\scriptscriptstyle|x-y|\geq d_\lambda}}} \hspace{-0.2cm} e^{-(|z-y|+|x-y|)/(2d')}\mathds{1}_{\scriptscriptstyle\theta_z \neq \theta_y} \Big)\cr
		&\leq \mathcal{O}\Big(\hspace{-0.2cm}\sum_{\substack{x,y,z\\ \scriptscriptstyle|x-y|\leq d_\lambda}} \hspace{-0.2cm}e^{-|z-y|/(2d')}\mathds{1}_{\scriptscriptstyle\theta_z \neq \theta_y} \left(e^{-\max(d_x,d_y)/(2d')}
		+e^{-2\Delta/\delta}\right)+\hspace{-0.2cm}\sum_{\substack{x,y,z\\ \scriptscriptstyle|x-y|\geq d_\lambda}}  \hspace{-0.2cm} e^{-(|z-y|+|x-y|)/(4d')}\mathds{1}_{\scriptscriptstyle\theta_z \neq \theta_y} \Big)\cr
		&\leq\mathcal{O}\Big(e^{-(L/2-2d_\lambda)/(4d')}\sum_{x,y,z}  e^{-(|z-y|+|y-x|)/(4d')}\mathds{1}_{\scriptscriptstyle\theta_z \neq \theta_y}+\sum_{z,y} e^{-|z-y|/(2d')-2\Delta/\delta}\mathds{1}_{\theta_z \neq \theta_y}(1+d_\lambda)\cr
		&\hspace{1cm}
		+e^{-d_\lambda/(8d')}\sum_{x,y,z} e^{-(|z-y|+|x-y|)/(8d')}\mathds{1}_{\theta_z \neq \theta_y} \Big)
\end{align}\vspace{-0.5cm}

We used the following properties. First, when $|x-y|\leq d_\lambda$ then 
$$e^{-(|z-y|+\max(d_x,d_y)/(2d'))/(4d')}\mathds{1}_{\theta_z \neq \theta_y}\leq e^{-(L-d_\lambda)/(4d')}\mathds{1}_{\theta_z \neq \theta_y}\leq e^{-(L-2d_\lambda)/(4d')}e^{-|x-y|/(4d')}\mathds{1}_{\theta_z \neq \theta_y}.$$ 
Second, because we are on a chain, we have $\sup_x\sum_{y,|x-y|\leq d_\lambda} = \sup_x N_{d_\lambda}(x) \leq (1+d_\lambda)$. Finally, we use that $\sup_x \sum_y e^{|x-y|/(16d')}=\mathcal{O}(1)$ and $\sup_{x, \theta(x)=1}=N=\mathcal{O}(1)$ together with the fact that $\theta_z\neq \theta_x$ imply that  $z$ and $y$ are on either side of the transition of $\theta$. Because this transition as been put at a distance $L/2$ of the nearest edge, this implies that

\begin{equation}
	||\{A,S\}||_1 =\mathcal{O}\Big(e^{-(L/2-2d_\lambda)/(4d')}+e^{-2\Delta/\delta}(1+d_\lambda)+e^{-d_\lambda/(8d')}\Big)
\end{equation}
If we then take $d_\lambda = L/6$, it implies:
\begin{equation}
	||\{A,S\}||_1=\mathcal{O}\left(e^{-2\Delta/\delta}+e^{-L/(48d')}\right)
\end{equation}
In order to prove the same result for $\norm{[1-S^2,\theta]}_1$, one need first to use:
\begin{align}
    \norm{[1-S^2,\theta]}_1&\leq \sum_{x,y} |(1-S^2)_{x,y} (\theta(x)-\theta(y))| \cr
    &\leq \sum_{x,y}\left| \min\left(e^{-\max(d_x,d_y)/(2d')} +e^{-2\Delta/\delta},e^{-|x-y|/(2d')}\right) \mathds{1}_{\theta_z \neq \theta_y}\right|
\end{align}
Then using exactly the same tricks as for $||\{A,S\}||_1$ (but in a simplified manner as we sum only on two and not three indices) one obtains that:

\begin{equation}
	\norm{[1-S^2,\theta]}_1=\mathcal{O}\left(e^{-2\Delta/\delta}+e^{-L/(48d')}\right)
\end{equation}

\appendix


\end{document}